\documentclass{aip-cp}

\usepackage[numbers]{natbib}
\usepackage{graphicx}
\usepackage{esvect}
\usepackage[utf8]{inputenc}

\begin{document}

\title{Liouville's Theorem from the Principle of Maximum Caliber in Phase Space}

\author[aff1]{Diego González\corref{cor1}}
\author[aff2]{Sergio Davis}
\eaddress{sdavis@cchen.cl}

\affil[aff1]{Grupo de Nanomateriales, Departamento de Física, Facultad de
Ciencias, Universidad de Chile.}
\affil[aff2]{Comisión Chilena de Energía Nuclear, Casilla 188-D, Santiago,
Chile.}
\corresp[cor1]{Corresponding author: dgonzalez@gnm.cl}

\maketitle

\begin{abstract}
One of the cornerstones in non--equilibrium statistical mechanics (NESM) is
Liouville's theorem, a differential equation for the phase space probability
$\rho(q,p; t)$. This is usually derived considering the flow in or out of a given surface for a
physical system (composed of atoms), via more or less heuristic arguments. 

In this work, we derive the Liouville equation as the partial differential
equation governing the dynamics of the time-dependent probability $\rho(q, p; t)$ of finding a 
``particle'' with Lagrangian $L(q, \dot{q}; t)$ in a specific point $(q, p)$ in
phase space at time $t$, with $p=\partial L/\partial \dot{q}$. This derivation depends only on 
considerations of inference over a space of continuous paths. Because of its
generality, our result is valid not only for ``physical'' systems but for any model 
depending on constrained information about position and velocity, such as time series.
\end{abstract}

\section{INTRODUCTION}

Equilibrium Statistical Mechanics was unified and shown to be an inference
procedure by E. T. Jaynes in 1957, who introduced the (now well-established) principle of Maximum
Entropy~\cite{Jaynes1957}. This principle provides a systematic procedure to
construct a probability distribution and, from it, to obtain macroscopic
properties of thermodynamic systems.

For non-equilibrium systems there are few exact results, most expressed in terms
of differential equations for time-dependent properties and probability
distributions. As a starting point in many derivations of classical statistical mechanics lies the Liouville equation 

\begin{equation}
\frac{d\rho}{dt} = \frac{\partial \rho}{\partial t} + \Big\{\rho, \mathcal{H}\Big\} = 0,
\end{equation}
which describes the time evolution of the probability density of microstates
with constant energy. Besides this fundamental equation, there is no unifying
axiomatic approach, in strong contrast with the equilibrium theory. 

As such unifying proposal, Jaynes in 1980 introduced a straightforward extension
of his Maximum Entropy principle, the \emph{Maximum Caliber}
Principle~\cite{Jaynes1980}, to dynamical systems. In this proposal, the path entropy or \emph{caliber} is maximized
subject to constraints to obtain the most unbiased path probability distribution.

We have previously~\cite{Gonzalez2014} shown that Maximum Caliber predicts
Newton's second law $F=m\cdot a$ for paths from informational constraints on
position and velocity, and recently~\cite{Davis2015} that a Boltzmann transport
equation arises naturally from the formalism in phase space for the same
constraints.

In this work we continue in the same line of reasoning, and show that the Liouville equation 
is the fundamental equation that describes the instantaneous probability distribution $\rho(q,p; t)$ in phase space 
in a problem of inference over an ensemble of trajectories $\bar{q}(t)$ and $\bar{p}(t)$, described by a
probability functional $P[\bar{q}, \bar{p}]$.

\section{The Maximum Caliber Principle}

The Maximum Caliber principle can be seen as a generalization of the Maximum
Entropy principle for assigning a probability distribution to an ensemble of
trajectories (or paths) given some information in the form of expectations. 

Consider a dynamical system described by a path $\bar{x}() \in \mathbb{X}$, where
$\mathbb{X}$ represents the space of continuous paths $\bar{x}(t)$ with $t \in
[0, \tau]$ and fixed boundary conditions $x(0)=x_0$ and $x(\tau)=x_\tau$. If the information 
we have about the system is given as the known value $G_0$ of the expectation of a functional $G[\bar{x}()]$, that is,

\begin{equation}
\big<G[\bar{x}()]\big>=\int_\mathbb{X} \mathcal{D}\bar{x} P[\bar{x}()] G[\bar{x}()] = G_0,
\label{functionalconst}
\end{equation}
then the most unbiased probability functional $P[\bar{x}()]$ compatible with this
constraint and the implicit constraint of normalization over $\mathbb{X}$,

\begin{equation}
\big<1\big>=\int_\mathbb{X} \mathcal{D}\bar{x} P[\bar{x}()] = 1,
\label{normalize}
\end{equation}
is the one that maximizes the \emph{caliber}

\begin{equation}
C[P] = -\int_\mathbb{X} \mathcal{D}\bar{x} P[\bar{x}()] \ln \frac{P[\bar{x}()]}{\pi[\bar{x}()]}.
\label{Caliber}
\end{equation}
which is in fact the Shannon-Jaynes entropy in path space with $\pi[\bar{x}()]$ the 
invariant measure of $\bar{x}$. The probability functional which maximizes $C[P]$ is of the form

\begin{equation}
P[\bar{x}()|G_0] = \pi[\bar{x}()]\frac{\exp(-\beta G[\bar{x}()])}{Z(\beta)},
\end{equation}
with $\beta$ a Lagrange multiplier, determined by the constraint equation

\begin{equation}
-\frac{\partial}{\partial \beta}\ln Z(\beta) = G_0.
\end{equation}
This is the approach most similar to maximum entropy. 
On the other hand, if the information we have on the system is given as an infinite set of 
instantaneous constraints 

\begin{equation}
\big<L[\bar{x}();t]\big>=L_0(t)
\label{functionconst}
\end{equation}
for each value of $t \in [0, \tau]$, maximization of $C[P]$ leads to the probability functional

\begin{equation}
P[\bar{x}()|L_0()] = \frac{\exp\left(-\int_0^\tau dt \beta(t) L[\bar{x}();t]\right)}{Z[\beta()]},
\end{equation}
where now the Lagrange multiplier $\beta()$ is a function of time, determined by 

\begin{equation}
-\frac{\delta}{\delta \beta(t)}\ln Z[\beta()] = L_0(t).
\end{equation}
This is equivalent to the first kind of constraint (Eq. \ref{functionalconst}) if we identify the functional $G$ as

\begin{equation}
\beta G[\bar{x}()]=\int_0^\tau dt\beta(t)L[\bar{x}();t].
\end{equation}

\subsection{Time-slicing of the path distribution in phase space and the continuity equation for probability}

Although the Maximum Caliber principle is promising as the definitive solution
for inference of time-independent systems (including non-equilibrium Statistical
Mechanics), it comes expressed in the language of paths $\bar{x}(t)$ and probability
functionals $P[\bar{x}()]$. In practice it is often needed to express the
information in terms of the instantaneous probability density of position
$\rho(x; t)$ and the joint probability distribution of position and velocity,
$\rho(x,\dot{x}; t)$. The procedure that discards information when going from the functional 
representation to the instantaneous representation we will call \emph{time-slicing}. The \emph{time-slicing
problem} is, then, to define the general procedure by which to obtain instantaneous probability 
densities for arbitrary quantities $G[\bar{x}(); t]$ at arbitrary times $t \in [0, \tau]$.

For a classical mechanical system with a well-defined Lagrangian $\mathcal{L}(q,\dot{q};t)$ a 
phase space can be constructed by defining the momentum $p$ as

\begin{equation}
p = \left(\frac{\partial \mathcal{L}}{\partial \dot{q}}\right).
\end{equation}

In the same way, each coordinate trajectory $\bar{q}(t)$ has an associated ``momentum trajectory'' 
$\bar{p}(t)$ obtained by applying the same rule for each time $t \in [0, \tau]$,

\begin{equation}
\bar{p}(t)=\left(\frac{\partial \mathcal{L}}{\partial \dot{q}}\right)\Big|_t.
\end{equation}

If we have a path space $\Gamma$ with fixed initial position and momentum, for instance,
$q(0)=q_0$ and $p(0)=p_0$, then the normalization condition for the probability functional is

\begin{equation}
\big <1\big> = \int_\Gamma D\bar{q} D\bar{p} P[\bar{q}(), \bar{p}()] = 1.
\label{normalize22}
\end{equation}

Now we want to determine the probability $\rho(q', p'; t)$ of being at a point $(q', p')$ in phase space at
a posterior time $t > 0$. This will be given by

\begin{equation}
\rho(q',p',t) = \big<\delta(\bar{q}(t)-q')\delta(\bar{p}(t)-p')\big>.
\label{slice}
\end{equation}
If we take a partial derivative with respect to $t$ of the above expression,
then we obtain 

\begin{equation}
\frac{\partial \rho(q', p', t)}{\partial t} + \frac{\partial}{\partial q'}\big<\delta(\bar{q}(t)-q') \delta(\bar{p}(t)-p')\frac{d\bar{q}}{dt}(t)\big> 
+ \frac{\partial}{\partial p'} \big < \delta(\bar{q}(t)-q') \delta(\bar{p}(t)-p')\frac{d\bar{p}}{dt}(t)\big> = 0,
\end{equation}
and, rewriting the expectations of the delta functions in the form

\begin{eqnarray}
\big<\delta(\bar{q}(t)-q') \delta(\bar{p}(t)-p')\frac{d\bar{q}}{dt}(t)\big> = \rho(q', p'; t)\big<\frac{d\bar{q}}{dt}(t)\big>_{c} \\
\big <\delta(\bar{q}(t)-q') \delta(\bar{p}(t)-p')\frac{d\bar{p}}{dt}(t)\big> = \rho(q', p'; t)\big<\frac{d\bar{p}}{dt}(t)\big>_{c},
\end{eqnarray}
where the subscripted expectation $\big<\cdot\big>_c$ denotes a time-sliced
expectation, conditional to $\bar{q}(t)=q'$ and $\bar{p}(t)=p'$, we find that the 
probability density $\rho$ for the phase space coordinate $(q', p')$ follows a continuity equation,

\begin{equation}
\frac{\partial \rho(q', p'; t)}{\partial t} +
\frac{\partial}{\partial q'}\left(\rho(q', p'; t)U_q(q', p'; t)\right) +
\frac{\partial}{\partial p'}\left(\rho(q', p'; t)U_p(q', p'; t)\right) = 0,
\label{eq_continuity}
\end{equation}
with current velocity components 

\begin{eqnarray}
U_q = \Big<\frac{d\bar{q}}{dt}(t)\Big>_c, \\
U_p = \Big<\frac{d\bar{p}}{dt}(t)\Big>_c.
\end{eqnarray}
In order to obtain a particular partial differential equation for $\rho$ we must
first determine the explicit form of this current velocity as a function of the 
phase space coordinate. In order to do this in the general case, we will need to
introduce some assumptions about the particular form of the probability functional 
$P[\bar{q}(),\bar{p}()]$, and this we will achieve by invoking the maximum caliber principle.

\section{Probability Functional for Paths in Phase Space}

If the classical action is constrained for a physical system by

\begin{equation}
\left<\int_0^\tau dt \mathcal{L}(\bar{q}(t), \frac{d\bar{q}}{dt}(t); t)\right> = \int \mathcal{D}q P[\bar{q}()] S[\bar{q}()] = S_0
\label{Action}
\end{equation}
then, according to the maximum caliber principle, the most unbiased probability
functional for the coordinate path $\bar{q}$ that agrees with that information will be given by

\begin{equation}
P[\bar{q}()|S_0] = \frac{\exp(-\beta \int_0^\tau dt \mathcal{L}(\bar{q}(t), \frac{d\bar{q}}{dt}(t); t)}{Z(\beta)},
\label{prob_q}
\end{equation}
with $\beta$ a Lagrange multiplier. From this we can construct the probability
functional for the phase space path $P[\bar{q}(), \bar{p}()]$ using the product rule of probability theory,

\begin{equation}
P[\bar{q}(), \bar{p}()|S_0] = P[\bar{q}()|S_0] P[\bar{p}()|\bar{q}(),S_0],
\label{prod}
\end{equation}
and the fact that complete knowledge of $\bar{q}()$ implies complete knowledge of
$\bar{p}()$, so $P[\bar{p}()|\bar{q}(),S_0]$ must be a functional Dirac's delta,

\begin{equation}
P[\bar{p}()|\bar{q}(),\mathcal{I}] = \delta\left[\frac{\partial L}{\partial \dot q} -
\bar{p}()\right] \approx \prod_i\delta\left(\frac{\partial L}{\partial \dot q(t_{i})} - p(t_{i})\right),
\label{delta_p}
\end{equation}
for any proposition $\mathcal{I}$ not in conflict with $\bar{q}()$. This strong
constraint, which fixes $\bar{p}(t)$ for every instant $t \in [0, \tau]$, can be translated into a single delta 
constraint, $P[\bar{p}()|\bar{q}()]=\delta(\epsilon_p[\bar{q}()])$, where we introduce a new
functional $\epsilon_p$,

\begin{equation}
\epsilon_p[\bar{q}()] = \int_0^\tau dt \left[\bar{p}(t)-\frac{\partial L}{\partial \dot q(t)}\right]^2.
\end{equation}

Now our phase space probability functional can be written as

\begin{equation}
P[\bar{q}(), \bar{p}()|S_0]
=\frac{1}{\eta(\beta)}\delta(\epsilon_p[\bar{q}()])\exp\left(-\beta\int_0^\tau dt
\mathcal{L}(\bar{q}(t), \frac{d\bar{q}}{dt}(t); t)\right)
\end{equation}
where $\eta$ is a normalization factor,

\begin{equation}
\eta(\beta) = \int \mathcal{D}\bar{q}\mathcal{D}\bar{p}\;
\delta(\epsilon_p[\bar{q}()])\exp\left(-\beta\int_0^\tau dt \mathcal{L}(\bar{q}(t),
\frac{d\bar{q}}{dt}(t); t)\right),
\end{equation}
and where the delta function on $\epsilon_p$ plays the role of a prior probability on phase space. 
The quadratic form of the integrand in $\epsilon_p$ makes it relatively easy to evaluate its functional derivatives. 
It only remains to write the exponent in terms of $\bar{q}(t)$ and $\bar{p}(t)$, and for this we use the Legendre transformation 

\begin{equation}
\mathcal{H}(q,p) = p\dot{q} - \mathcal{L}(q, \dot{q}; t)
\end{equation}
and insert the Hamiltonian $\mathcal{H}$, leading to

\begin{equation}
P[\bar{q}(), \bar{p}()|S_0]
=\frac{1}{\eta(\beta)}\delta(\epsilon_p[\bar{q}()])\exp\left(-\beta\int_0^\tau dt \left[ \bar{p}(t)\frac{d\bar{q}}{dt}(t)-\mathcal{H}(\bar{q}(t), \bar{p}(t), t)\right]\right).
\label{eq_prob_phasespace}
\end{equation}

We will extract information about the current velocity in the continuity
equation (Eq. \ref{eq_continuity}) by using a recently derived general identity
between expectations known as the conjugate variables theorem~\cite{Davis2012, Davis2015}.

\section{Functional form of the Conjugate Variables Theorem}

The conjugate variables theorem~\cite{Davis2012} (CVT) for $N-$dimensional state space is the identity

\begin{equation}
\Big<\frac{\partial \omega(\vec x)}{\partial x_i}\Big> + \Big<\omega(\vec x)\frac{\partial}{\partial x_i} \ln \rho(\vec x)\Big> = 0.
\end{equation}
where $\omega(x)$ is an arbitrary, differentiable function of the state $x$, and
$\rho(x)$ is the time-independent probability distribution of $x$. Its natural extension to probability functionals is 

\begin{equation}
\Big<\frac{\delta W[x()]}{\delta x(t)}\Big> + \Big<W[x()]\frac{\delta}{\delta x(t)}\ln P[x()]\Big> = 0,
\end{equation}
as put forward in Ref. \cite{Davis2015}. For our probability functional in phase space (Eq. \ref{eq_prob_phasespace}), the CVT gives us two independent identities,

\begin{eqnarray}
\Big<\frac{\delta W[\bar{q}(),\bar{p}()]}{\delta \bar{q}(t)}\Big> = -\beta\Big<W[\bar{q}(),\bar{p}()]\left(\frac{d\bar{p}}{dt}(t)+\frac{\partial \mathcal{H}}{\partial q}\right)\Big> - \Big<W[\bar{q}(),\bar{p}()]\frac{\delta}{\delta \bar{q}(t)}\ln \delta(\epsilon_p[\bar{q}()])\Big>, \nonumber \\
\Big<\frac{\delta W[\bar{q}(),\bar{p}()]}{\delta \bar{p}(t)}\Big> = \beta\Big<W[\bar{q}(),\bar{p}()]\left(\frac{d\bar{q}}{dt}(t)-\frac{\partial \mathcal{H}}{\partial p}\right)\Big> - \Big<W[\bar{q}(),\bar{p}()]\frac{\delta}{\delta \bar{p}(t)}\ln \delta(\epsilon_p[\bar{q}()])\Big>.
\label{eq_cvt_phase_space}
\end{eqnarray}

Note that in the expressions above we use the $\delta$ symbol to denote both the functional derivative $\delta/\delta \bar{q}$ and 
Dirac's delta function $\delta(\cdot)$, each use we hope is clear in context. The functional derivatives of $\delta(\epsilon_p[\bar{q}()])$ 
involve the functional derivatives of $\epsilon_p[\bar{q}()]$,

\begin{eqnarray}
\frac{\delta}{\delta \bar{q}(t)}\epsilon_p[\bar{q}()] = -2\left[\bar{p}(t)-\frac{\partial
\mathcal{L}}{\partial \dot{q}}\right]\left(\frac{\partial^2
\mathcal{L}}{\partial \dot{q}\partial q}-\frac{d}{dt}\left(\frac{\partial^2
\mathcal{L}}{\partial \dot{q}^2}\right)\right), \\
\frac{\delta}{\delta \bar{p}(t)}\epsilon_p[\bar{q}()] =
2\left[\bar{p}(t)-\frac{\partial \mathcal{L}}{\partial \dot{q}}\right],
\end{eqnarray}
through the chain rule, and these derivatives are zero when evaluated on the condition itself. Therefore, the second
terms in the right-hand side of Eqs. \ref{eq_cvt_phase_space} both vanish for
any $W$. The CVT reduces then to 

\begin{eqnarray}
\Big<\frac{\delta W[\bar{q}(),\bar{p}()]}{\delta \bar{q}(t)}\Big> =
-\beta\Big<W[\bar{q}(),\bar{p}()]\left(\frac{d\bar{p}}{dt}(t)+\frac{\partial \mathcal{H}}{\partial
q}\right)\Big>, \nonumber \\
\Big<\frac{\delta W[\bar{q}(),\bar{p}()]}{\delta \bar{p}(t)}\Big> =
\beta\Big<W[\bar{q}(),\bar{p}()]\left(\frac{d\bar{q}}{dt}(t)-\frac{\partial \mathcal{H}}{\partial p}\right)\Big>.
\label{eq_cvt_func_phasespace}
\end{eqnarray}
As an example of the use of this general identity, we take the simplest
non-trivial choice $W=1$, and readily obtain Hamilton's equations in expectation,

\begin{eqnarray}
\Big<\frac{d\bar{q}}{dt}(t)\Big> = \Big<\left(\frac{\partial H}{\partial p}\right)\Big|_t\Big>, \\
\Big<\frac{d\bar{p}}{dt}(t)\Big> = \Big<\left(\frac{\partial H}{\partial q}\right)\Big|_t\Big>,
\end{eqnarray}
which are in agreement with Newton's second law in the particular case obtained
in Ref.~\cite{Gonzalez2014} from Maximum Caliber.

\section{Derivation of the Liouville equation}

Now we are equipped to compute the components of the current velocity $U_q$ and
$U_p$ in the continuity equation (Eq. \ref{eq_continuity}). To construct these time-sliced expectations we take 
Eq. \ref{eq_cvt_func_phasespace} and choose $W=\delta(\bar{q}(t)-q')\delta(\bar{p}(t)-p')$. After some more or less 
straightforward computation we obtain 

\begin{eqnarray}
\Big<\frac{d\bar{q}}{dt}(t)\Big>_c = U_q = \frac{1}{\beta}\frac{\partial}{\partial p'}\ln \rho + \frac{\partial H}{\partial p'}, \nonumber \\
\Big<\frac{d\bar{p}}{dt}(t)\Big>_c = U_p = -\frac{1}{\beta}\frac{\partial}{\partial q'}\ln \rho - \frac{\partial H}{\partial q'}.
\end{eqnarray}
We are ready to place these expressions into the continuity equation (Eq. \ref{eq_continuity}) from which 
we readily find that the divergence of the current velocity is zero:

\begin{equation}
\frac{\partial}{\partial q'}U_q + \frac{\partial}{\partial p'}U_p = 0.
\end{equation}
Therefore the continuity equation becomes the Liouville equation:

\begin{equation}
\frac{\partial \rho}{\partial t} + \Big\{\rho, H\Big\} = 0.
\label{louville}
\end{equation}

\section{CONCLUSIONS}

We have obtained the Liouville equation as a direct consequence of the Maximum
Caliber formalism for any system with classical action given by the integral of a
Lagrangian function dependent on position and velocity. The derivation is
self--contained: it does not assume anything besides the validity of the Maximum
Caliber formalism and the Legendre transformation. This makes the use of phase
space ideas in inference for dynamical systems natural and independent of
physical considerations, being for instance suited to the study of time series.

\section{ACKNOWLEDGMENTS}

The authors acknowledge support from FONDECYT grant 1140514. DG acknowledges
funding from CONICYT PhD fellowship 21140914.


\bibliographystyle{aipnum-cp}
\bibliography{cvt}

\end{document}